\documentclass[twocolumn,aps,pra,showpacs]{revtex4}
\usepackage{epsfig}
\begin{document}
\title{Partial deterministic non-demolition Bell measurement}
\author{Ladislav Mi\v{s}ta Jr. and Radim Filip}
\affiliation{Department of Optics, Palack\' y University,\\
17. listopadu 50,  772~07 Olomouc, \\ Czech Republic}
\date{\today}
\begin{abstract}
We propose the scheme implementing partial deterministic non-demolition Bell 
measurement. When it is used in quantum teleportation the information about 
an unknown input state is optimally distributed among three outgoing qubits. 
The optimality means that the output fidelities saturate the asymmetric cloning
inequality. The flow of information among the qubits is controlled by the 
preparation of a pair of ancillary qubits used in the Bell measurement. 
It is also demonstrated that the measurement is optimal two-qubit operation
in the sense of the trade-off between the state disturbance and
the information gain.  

\end{abstract}

\maketitle

\begin{figure}
\centerline{\psfig{width=6.5cm,angle=0,file=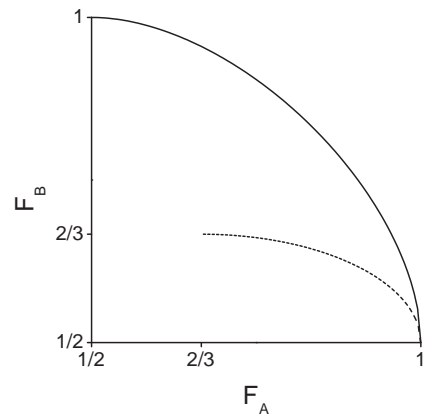}}
\vspace{-0.5cm}
\caption{Bounds for the fidelities $F_{A}$ and $F_{B}$ in the partial classical teleportation 
(dashed curve) and the partial quantum teleportation (solid curve).}
\label{fig0}
\end{figure}

Quantum entanglement makes it possible to solve some quantum communication
tasks better. This situation is encountered for instance in the 
teleportation of quantum states. In this task, one party called Alice, is 
requested to transmit an {\it unknown} pure state $|\psi\rangle_{A}$ of a
qubit $A$ to the second party called Bob without sending the qubit itself. 
There are two ways how she can accomplish the task. First, she can perform 
an optimal measurement on the state and send the results to Bob. After receiving
the classical information from Alice Bob is able to estimate the
input state on another qubit $B$. Such a protocol, sometimes called  
classical teleportation, however allows Alice to transmit the input 
state only approximately. Using the mean fidelity as a characterization of 
the quality of the transfer the fidelity $F_{B}$ of Bob's estimated state 
can be at most $F_{B}=2/3$ \cite{Massar_95}. 
In addition, the measurement on Alice's side disturbs the original qubit $A$ in 
such a way that it does not carry any further information on the input 
state and therefore the mean fidelity of Alice's output state $F_{A}$ cannot
exceed $F_{A}=2/3$ \cite{Banaszek_01}. In what follows we denote the fidelities
$F_{A}$ and $F_{B}$ as operation and teleportation fidelity, respectively. 
It is possible to consider a more general classical 
teleportation in which the information on the input state $|\psi\rangle_{A}$ 
is distributed asymmetrically between the qubits $A$ and $B$. This regime 
is achieved if Alice performs less demolishing measurement after
which her qubit carries more information on the input state and
therefore $F_{A}>2/3$ while she gains less information and 
therefore $F_{B}<2/3$. Quantum mechanics imposes a fundamental 
constraint on the fidelities $F_{A}$ and $F_{B}$ in the {\it partial} classical
teleportation (PCT) in the form of the following inequality \cite{Banaszek_01}:
\begin{eqnarray}\label{Banaszek}
\sqrt{F_{A}-1/3}\leq\sqrt{F_{B}-1/3}+\sqrt{2/3-F_{B}}.  
\end{eqnarray}
The equality in inequality (\ref{Banaszek}) corresponds to an {\it optimal} PCT 
in the sense, that for a given operation fidelity $F_{A}$ a larger teleportation 
fidelity $F_{B}$ cannot be obtained. The optimal fidelities lie on the fragment 
of the ellipse depicted by the dashed curve in Fig.~\ref{fig0} \cite{Banaszek_01}.

Better and even perfect transfer of an unknown state can be achieved using 
another procedure now known as quantum teleportation \cite{Bennett_93}. 
In this protocol, Alice and Bob share a maximally entangled state of 
two qubits $a$ and $B$. Alice then performs a specific joint measurement, 
called Bell measurement, on the input qubit $A$ whose state 
is to be transferred and on a qubit $a$. As a result, Bob's qubit $B$ 
collapses into the state that can be brought into the original state 
by an appropriate unitary transformation. Thus, at the output of the 
quantum teleportation the qubit $B$ carries a perfect replica of the original 
state and therefore the teleportation fidelity is $F_{B}=1$ while the 
qubit $A$ has to carry no information on the input state and therefore 
the operation fidelity is $F_{A}=1/2$.

Obviously, Bell measurement is the crucial ingredient of quantum 
teleportation. This is the two-qubit projection measurement 
discriminating perfectly among four orthogonal maximally 
entangled Bell states. Motivated by the classical partial 
teleportation we can think about a {\it partial} non-demolition Bell 
measurement that discriminates only 
partially among the Bell states and simultaneously preserves all Bell
states. Such a measurement would realize just a {\it partial} quantum 
teleportation (PQT) of an unknown input state onto qubit $B$ while 
preserving some information on the state in the input qubit $A$.
In addition, in the PQT one can expect a larger teleportation fidelity
$F_{B}$ for a given operation fidelity $F_{A}$ than in the PCT.
Also for PQT quantum mechanics sets a nontrivial bound between the fidelities 
$F_{A}$ and $F_{B}$ in the form of the cloning inequality 
\cite{Niu_98}
\begin{equation}\label{inequality}
(1-F_{A})(1-F_{B})\geq[1/2-(1-F_{A})-(1-F_{B})]^{2},  
\end{equation}
where the equality corresponds to an {\it optimal} PQT in the sense that a 
larger teleportation fidelity $F_{B}$ cannot be obtained for a given operation 
fidelity $F_{A}$. The bound for the fidelities is again given by the fragment 
of the ellipse (\ref{inequality}) depicted by the solid curve in Fig.~\ref{fig0}. 
The Fig.~\ref{fig0} reveals that the PQT really provides a larger 
teleportation fidelity $F_{B}$ for a given operation fidelity $F_{A}$ 
than the PCT. 

In this article we construct such a partial non-demolition Bell measurement (PNBM) 
for qubits that allows optimal PQT. The measurement rests on the scheme for 
complete non-demolition Bell measurement proposed in \cite{Guo_01}. The novel 
point is that two qubits used as ancillas in the measurement are prepared 
in a specific partially entangled state depending on a
single parameter whose change allows to continuously control the degree 
of discrimination of the Bell states. Further we show, that the
PNBM is also an optimal two-qubit 
operation providing maximum estimation fidelity ${\cal F}_{B}$ for a 
given operation fidelity ${\cal F}_{A}$, i. e. the fidelities saturate the 
two-qubit analogy of the inequality (\ref{Banaszek}) \cite{Banaszek_01}
\begin{eqnarray}\label{Banaszek2}
\sqrt{{\cal F}_{A}-1/5}\leq\sqrt{{\cal F_{B}}-1/5}+\sqrt{3\left(2/5-{\cal F_{B}}\right)}.  
\end{eqnarray}

Note, that the PQT differs from the telecloning or cloning 
``at a distance'' \cite{Murao_99} as it realizes rather the ``nonlocal'' 
cloning as the qubits carrying the information on the input state emerge 
at different possibly distant locations. Recently, a conditional 
scheme realizing optimal PQT was proposed in \cite{Filip_04b} based 
on the partial discrimination of a singlet state the degree of which
being controlled by the splitting ratio of the beam splitter used for 
the discrimination. Here, we concentrate on the {\it deterministic} scheme. 
The performance of our scheme relies on the measurement discriminating 
partially among all four Bell states. In our scheme, the degree of 
discrimination is controlled by the preparation of 
the pair of qubits used as ancillas in the measurement.

The paper is structured as follows. In Section~\ref{sec_qubits} 
a PNBM for qubits is designed and an optimal PQT protocol is constructed. 
Section~\ref{sec_interpret} contains interpretation of the PNBM. 
Section~\ref{sec_CV} deals with continuous-variable (CV) 
partial teleportation of coherent states. Section~\ref{sec_conclusion} 
contains conclusion.
\section{Partial non-demolition Bell measurement for qubits}\label{sec_qubits}

Let us consider the following task. Alice has at her disposal
a qubit $A$ in an {\it unknown} pure normalized state $|\psi\rangle_{A}=a|0\rangle_{A}+b|1\rangle_{A}$ 
($|0\rangle$ and $|1\rangle$ denote the eigenvectors of the Pauli diagonal matrix 
$\sigma_{3}=\mbox{diag}(1,-1)$ corresponding to the eigenvalues $+1$ and $-1$, respectively) 
and she would like to {\it optimally} partially teleport the state to Bob. 
According to the definition in introduction optimal PQT means that after the teleportation 
Alice and Bob each hold an imperfect copy $\rho_{A}$ and $\rho_{B}$, respectively, of 
the original state whose fidelities $F_{A}$ and $F_{B}$ saturate the inequality (\ref{inequality}). 
\begin{figure}
\centerline{\psfig{width=6.5cm,angle=270,file=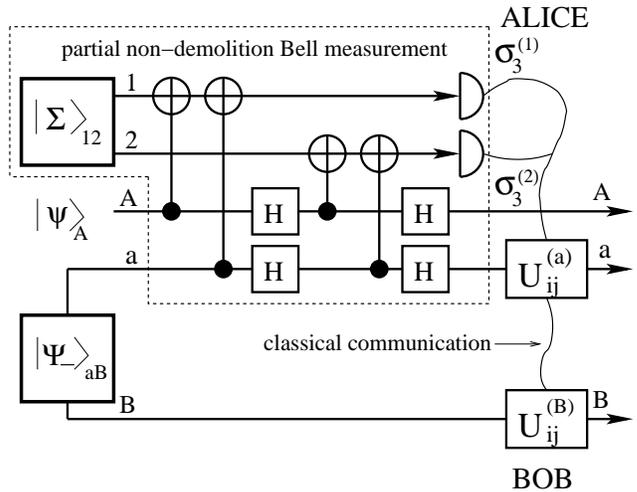}}
\caption{Schematic of the optimal partial quantum teleportation. 
The dashed box contains logical network for the
partial non-demolition Bell measurement. The lines connecting the 
qubits represent CNOT gates with control qubit indicated by full circle 
and target qubit by empty circle; $H$ stands for the Hadamard gate.  
The unitary transformations converting the qubits $B$ and $a$ 
into the approximate replicas of the state $|\psi\rangle$ and 
$|\psi_\perp\rangle$ are denoted as $U_{ij}^{(B)}$ and $U_{ij}^{(a)}$, 
respectively. See text for details.}
\label{fig1}
\end{figure}

Alice can perform the task via the teleportation scheme depicted in Fig.~\ref{fig1}. 
For this purpose she needs four CNOT gates, four Hadamard gates, a pair of properly 
prepared auxiliary qubits and a shared entangled state. Initially, Alice and Bob 
share a pair of qubits $a$ and $B$ prepared in a singlet state 
$|\Psi_{-}\rangle_{aB}=(|01\rangle_{aB}-|10\rangle_{aB})/\sqrt{2}$.
The key ingredient of the protocol is the PNBM that is performed by Alice. 
This is in fact the perfect non-demolition Bell measurement \cite{Guo_01} 
on the qubits $A$ and $a$ in which the ancillary qubits $1$ and $2$ are 
prepared in the state of the form 
\begin{equation}\label{ancillas}
|\Sigma\rangle_{12}=\alpha|00\rangle_{12}+\beta|++\rangle_{12},
\end{equation}
where $\alpha,\beta\geq 0$, $\alpha^{2}+\alpha\beta+\beta^{2}=1$ and 
$|\pm\rangle=(|0\rangle\pm|1\rangle)\sqrt{2}$. For $\alpha\beta\ne0$ the state 
(\ref{ancillas}) is partially entangled. This follows from the fact that the 
purity $P_{1}=\mbox{Tr}(\rho_{1}^{2})=1-\alpha^{2}\beta^{2}/2$ of the reduced 
density matrix $\rho_{1}=\mbox{Tr}_{2}(|\Sigma\rangle_{1212}\langle\Sigma|)$ of the qubit $1$
is less than one for $\alpha\beta\ne0$. Further, since $\alpha\beta\leq 1/3$, 
where the equality holds for $\alpha=\beta=1/\sqrt{3}$, $P_{1}\geq 17/18$ for any 
$\alpha$ and $\beta$ and therefore the state (\ref{ancillas}) is never entangled 
maximally the largest amount of entanglement being achieved for $\alpha=\beta=1/\sqrt{3}$. 

The logical network in the PNBM (see dashed box in Fig.~\ref{fig1}) transforms 
the input state $|\psi\rangle_{A}|\Psi_{-}\rangle_{aB}|\Sigma\rangle_{12}$ to 
\begin{equation}\label{output}
\frac{\alpha}{2}\sum_{i,j=0}^{1}|ij\rangle_{12}U_{ij}|\Psi_{-}\rangle_{Aa}|\psi\rangle_{B}
+\beta|++\rangle_{12}|\psi\rangle_{A}|\Psi_{-}\rangle_{aB}.
\end{equation}
Here $U_{ij}=U_{ij}^{(a)}\otimes U_{ij}^{(B)}$ is the product of local unitary 
transformations on qubits $a$ and $B$, where $U_{00}^{(l)}=\sigma_{2}^{(l)}$, 
$U_{01}^{(l)}=\sigma_{1}^{(l)}$, $U_{10}^{(l)}=\sigma_{3}^{(l)}$, $U_{11}^{(a)}=-I^{(a)}$
and $U_{11}^{(B)}=I^{(B)}$, where $\sigma_{k}^{(l)}$, $k=1,2,3$ and $I^{(l)}$ are 
standard Pauli matrices and the identity matrix, respectively, of the qubit $l$.   
Measuring the ancillary qubits $1$ and $2$ in the basis $|0\rangle$, $|1\rangle$ one 
finds with probability $1/4$ one of four results $00,01,10$ and $11$
where the first (second) digit is the result of the measurement on the
qubit $1(2)$. Communicating the result to Bob via classical channel, performing 
the corresponding unitary transformations $\left({U_{ij}^{(l)}}\right)^{-1}={U_{ij}^{(l)}}$, $l=a,B$ on 
qubits $a$ and $B$, and taking into account the formula 
$U_{ij}|\Psi_{-}\rangle_{aB}=-|\Psi_{-}\rangle_{aB}$ we find that Alice and Bob in 
each run of the teleportation establish between themselves a normalized state of three qubits 
$A,a$ and $B$ of the form
\begin{equation}\label{final}
|\Phi\rangle=\alpha|\Psi_{-}\rangle_{Aa}|\psi\rangle_{B}-\beta|\psi\rangle_{A}
|\Psi_{-}\rangle_{aB}.
\end{equation}
The interpretation of the resulting state is straightforward. If $\alpha=0$, i. e. the ancillary 
qubits are prepared in the separable state $|++\rangle_{12}$ and therefore the PNBM does 
not discriminate among the Bell states, the input state is completely preserved in the original 
qubit. If, on the other hand, $\beta=0$, i. e. the ancillary qubits are prepared in the separable 
state $|00\rangle_{12}$ therefore the PNBM perfectly discriminates among the Bell states, the 
input state $|\psi\rangle_{A}$ is completely teleported to Bob. For $\alpha\beta\ne 0$, 
i. e. for partially entangled ancillary qubits when PNBM discriminates partially among the Bell 
states, the state (\ref{final}) represents a coherent superposition of two previous possibilities 
the probability amplitudes of which being controlled by the choice of the parameter $\alpha$ 
in the state (\ref{ancillas}). In other words, this intermediate case corresponds to the 
{\it partial} teleportation of the qubit $A$. In what follows we show that the quantum 
teleportation scheme is {\it optimal} in the sense of the inequality (\ref{inequality}).

Tracing the state (\ref{final}) over two qubits one finds the remaining qubit $i$ in the mixed state 
\begin{equation}\label{reduced}
\rho_{i}=F_{i}|\psi\rangle_{ii}\langle\psi|+(1-F_{i})|\psi_\perp\rangle_{ii}\langle\psi_\perp|,
\quad i=A,B,a,   
\end{equation}
where $|\psi_\perp\rangle=b^{\ast}|0\rangle-a^{\ast}|1\rangle$ is the state orthogonal 
to the input state $|\psi\rangle$ and
\begin{equation}\label{fidelities}
F_{A}=1-\frac{\alpha^{2}}{2},\quad F_{B}=1-\frac{\beta^{2}}{2},\quad 
F_{a}=\frac{\alpha^{2}+\beta^{2}}{2}.
\end{equation}
Since the inequality (\ref{inequality}) becomes equality for the latter fidelities  
our scheme really realizes an {\it optimal} PQT. Viewed from the point of view of 
quantum cloning, our scheme realizes a measurement induced and entanglement assisted 
optimal $1\rightarrow 2$ universal asymmetric quantum cloning machine \cite{Niu_98}.
In particular, if the superposition in the state (\ref{ancillas}) is 
balanced with $\alpha=\beta=1/\sqrt{3}$, then the information is distributed symmetrically between 
the qubits $A$ and $B$ with the fidelities $F_{A}=F_{B}=5/6$. In the language of quantum 
cloning this regime corresponds to an {\it optimal} symmetric universal quantum cloning 
\cite{Buzek_96}. Moreover, the third qubit $a$ leaves the teleportation in the state 
$\rho_{a}$ for which the fidelity $F_{a\perp}={}_{a}\langle\psi_\perp|\rho_{a}|\psi_\perp\rangle_{a}$ 
attains its maximum possible value $F_{a\perp}=2/3$. Therefore, our scheme also realizes deterministically 
the {\it optimal} universal NOT gate \cite{Buzek_99}.
\begin{figure}
\centerline{\psfig{width=6.0cm,angle=0,file=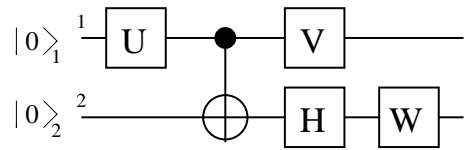}}
\caption{Logical network for preparation of the
state $|\Sigma\rangle_{12}$. The line connecting the qubits represents 
CNOT gate with control qubit indicated by full circle and target qubit 
by empty circle. $U$, $V$, are rotations, $W$ is the orthogonal transformation, 
and $H$ is the Hadamard gate.}
\label{fig2}
\end{figure}

It remains to prepare the state of ancillas $|\Sigma\rangle_{12}$.
This can be done with the help of a single CNOT gate and four local
unitary transformations as is depicted in Fig.~\ref{fig2}. The
transformations $U$ and $V$ are represented in the basis 
$|0\rangle$, $|1\rangle$ by the rotations $U$ and $V$ with elements
\begin{eqnarray*}
U_{11}&=&U_{22}=(\alpha+\beta+\sqrt{\alpha^{2}+\beta^{2}})/2,\\
U_{12}&=&-U_{21}=-(\alpha+\beta-\sqrt{\alpha^{2}+\beta^{2}})/2,\\ 
V_{11}&=&V_{22}=(\alpha+\sqrt{\alpha^{2}+\beta^{2}})/K,\\
V_{12}&=&-V_{21}=-\beta/K, 
\end{eqnarray*}
where $K=\sqrt{2(\alpha^{2}+\beta^{2}+\alpha\sqrt{\alpha^{2}+\beta^{2}})}$, $H$ is the 
Hadamard gate and $W$ is represented in the basis $|\pm\rangle$ by the orthogonal matrix 
$W$ with elements
\begin{eqnarray*}
W_{11}&=&\frac{\sqrt{2}U_{11}}{K},\quad W_{12}=\frac{{\beta}(\sqrt{\alpha^{2}+\beta^{2}}-\beta)}
{\sqrt{2}KU_{21}},\\ 
W_{21}&=&\frac{\alpha(\sqrt{\alpha^{2}+\beta^{2}}+\alpha)}{\sqrt{2}KU_{11}},\quad
W_{22}=-\frac{\alpha\beta}{\sqrt{2}KU_{21}}. 
\end{eqnarray*}
\section{Interpretation of the partial Bell measurement}\label{sec_interpret}

As it was already mentioned the standard Bell measurement is a projection 
measurement on a system of two qubits $A$ and $a$ discriminating perfectly among 
four orthogonal Bell states. Consequently, the measurement provides maximum 
information on an unknown pure two-qubit state and therefore the mean 
fidelity of the estimated state is ${\cal F}_{B}=2/5$ \cite{Bruss_99} while disturbing 
the input state in such a way that it does not carry any more information on 
the state and therefore the mean fidelity of the output state is also
${\cal F}_{A}=2/5$ \cite{Banaszek_01}. On the other hand, the partial Bell measurement 
considered in the preceding section discriminates among the Bell states only 
partially thus providing less information on the input state of qubits 
$A$ and $a$ while preserving larger information in the outgoing qubits. 
As the degree of the Bell state discrimination can be controlled by the
parameter $\alpha$ in the ancillary state (\ref{ancillas}), one can increase continuously 
the information gain at the expense of a larger disturbance of the measured system. 

To express this behaviour quantitatively let us assume the ancillary qubits $1$ and $2$ 
to be again prepared in the state (\ref{ancillas}). For such a state the  
measurement scheme in the dashed box in Fig.~\ref{fig1} realizes a two-qubit quantum 
operation on qubits $A$ and $a$ described by the set of four operators 
$A_{k}$, $k=1,2,\ldots,4$ where we have identified the indices $1,2,3$ and 
$4$ with the measurement outcomes $00,01,10$ and $11$. In the Bell basis 
$|\Phi_{1,2}\rangle=(|00\rangle\pm|11\rangle)\sqrt{2}$, 
$|\Phi_{3,4}\rangle=(|01\rangle\pm|10\rangle)\sqrt{2}$ the operators are represented 
by the diagonal matrices
\begin{eqnarray}\label{POVM}
A_{1}&=&\mbox{diag}(\alpha+\beta/2,\beta/2,\beta/2,\beta/2),\nonumber\\
A_{2}&=&\mbox{diag}(\beta/2,\alpha+\beta/2,\beta/2,\beta/2),\nonumber\\
A_{3}&=&\mbox{diag}(\beta/2,\beta/2,\alpha+\beta/2,\beta/2),\nonumber\\
A_{4}&=&\mbox{diag}(\beta/2,\beta/2,\beta/2,\alpha+\beta/2).
\end{eqnarray}
Consequently, the PNBM preserves the Bell states and therefore
it really realizes a non-demolition measurement of these states. Further, 
the operators (\ref{POVM}) satisfy the completeness relation 
$\sum_{i=1}^{4}A_{i}^{\dag}A_{i}=\openone$ due to the normalization 
condition $\alpha^{2}+\alpha\beta+\beta^{2}=1$ and hence the PNBM performs 
partial discrimination of the Bell states deterministically. 
Restricting ourselves to pure two-qubit input states, the information 
gain in the PNBM can be characterized by the mean estimation 
fidelity ${\cal F}_{B}$ while the disturbance of the state caused by the measurement 
can be characterized by the mean operation fidelity ${\cal F}_{A}$ \cite{Banaszek_01}. 
The fidelities can be calculated from the formulas \cite{Banaszek_01}
\begin{eqnarray}\label{formula2}
{\cal F}_{A}=\frac{\left(4+\sum_{i=1}^{4}|{\rm Tr}A_{i}|^2\right)}{20},\quad 
{\cal F}_{B}=\frac{\left(4+\sum_{i=1}^{4}\lambda_{i}\right)}{20},\nonumber\\ 
\end{eqnarray}
where $\lambda_{i}$ is the maximum eigenvalue of the matrix  $A_{i}^{\dag}A_{i}$. On inserting
the Eqs.~(\ref{POVM}) into the formulas (\ref{formula2}) we finally arrive at the fidelities 
of the PNBM 
\begin{eqnarray}\label{formula3}
{\cal F}_{A}=\frac{1+\left(\alpha+2\beta\right)^2}{5},\quad {\cal F}_{B}=
\frac{1+\left(\alpha+\beta/2\right)^2}{5}. 
\end{eqnarray}
Now an interesting question arises what is the trade-off between the obtained fidelities 
(\ref{formula3}). Substituting the fidelities (\ref{formula3}) into the inequality
(\ref{Banaszek2}) one finds that they saturate the inequality.
Hence, the PNBM is an {\it optimal} two-qubit quantum operation in the sense that 
for a given operation fidelity ${\cal F}_{A}$ a larger estimation 
fidelity ${\cal F}_{B}$ cannot be obtained. Therefore, the measurement 
allows an optimal PCT of two-qubit states. Moreover, as the same
optimal distribution of information can be achieved in quantum
teleportation with nonmaximally entangled states \cite{Banaszek_01}, 
our scheme represents a local counterpart to the quantum teleportation 
based on nonmaximally entangled states.
\section{Partial teleportation for continuous variables}\label{sec_CV}

Let us now study the extension of the optimal PQT to continuous-variables 
(CVs). In this case the five relevant qubits are replaced with the
five modes of electromagnetic field described by the quadrature 
operators $x_{i}$, $p_{i}$, $i=A,B,a,1,2$ with the commutators 
$[x_{j},p_{k}]=i\delta_{jk}$. The state of the
mode $A$ that is to be teleported is an {\it unknown} coherent state 
$|\alpha\rangle_{A}$. For CVs the optimal PQT means that after the 
teleportation Alice and Bob hold imperfect copies $\rho_{A}$ and 
$\rho_{B}$, respectively, of the original coherent 
state having the following fidelities 
$F_{i}=_{i}\!\!\langle\alpha|\rho_{i}|\alpha\rangle_{i}$, $i=A,B$: 
\begin{equation}\label{asymF}
F_{A}=\frac{2}{2+e^{2\gamma}},\quad F_{B}=\frac{2}{2+e^{-2\gamma}}, 
\end{equation}
where $\gamma$ is the asymmetry parameter \cite{Fiurasek_01}. 

The CV extension of the qubit protocol presented in Sec.~\ref{sec_qubits} 
is as follows. Initially, Alice and Bob share a two-mode squeezed vacuum 
state of modes $a$ and $B$ produced in the non-degenerate optical parametric
amplifier (NOPA)
\begin{equation}\label{NOPA}
|{\rm NOPA}\rangle_{aB}=\sqrt{1-\lambda^{2}}\sum_{n=0}^{\infty}\lambda^{n}|nn\rangle_{aB},
\end{equation}
where $\lambda=\tanh r$, $r$ is the squeezing parameter. Alice then performs 
the {\it CV partial non-demolition Bell measurement} on the mode $A$ whose 
state is to be teleported and on the mode $a$ of the NOPA state. Recently, the
measurement was used for partial reversion of the CV cloning
\cite{Filip_04a}. The measurement exploits two ancillary vacuum modes 
$1$, $2$ and a sequence of four CV CNOT gates followed by the homodyne 
detections of the $x$ quadrature on mode $1$ and $p$ quadrature on mode $2$ 
\cite{Filip_04a}. The CNOT gates are realized by the quantum non-demolition 
(QND) interaction
\begin{equation}\label{QND}
x_{j}'=x_{j},\quad x_{k}'=x_{k}+\kappa x_{j},\quad p_{j}'=p_{j}-\kappa p_{k},\quad p_{k}'=p_{k},
\nonumber\\
\end{equation}
where $j(k)$ is the control (target) mode and $\kappa$ is the coupling constant. By the 
measurement Alice detects the quadratures $x_{u}=x_{1}-\kappa(x_{A}-x_{a})$ and 
$p_{v}=p_{2}+\kappa(p_{A}+p_{a})$ for which she obtains two classical results 
$\bar x_{u}$ and $\bar p_{v}$. The quadratures of mode $A$ transform as 
\begin{eqnarray}\label{cA}
x_{A,{\rm out}}&=&x_{A}-\kappa x_{2},\quad p_{A,{\rm out}}=p_{A}+\kappa p_{1},
\end{eqnarray}
while the quadratures of the remaining two modes $a$ and $B$ can be written as
\begin{eqnarray}\label{acB}
x_{a}'&=&x_{A}-\frac{1}{\kappa}x_{1}-\kappa x_{2}+\frac{1}{\kappa}x_{u},\nonumber\\
p_{a}'&=&-p_{A}-\kappa p_{1}-\frac{1}{\kappa}p_{2}+\frac{1}{\kappa}p_{v},\nonumber\\
x_{B}&=&x_{A}-(x_{a}-x_{B})-\frac{1}{\kappa}x_{1}+\frac{1}{\kappa}x_{u},\nonumber\\
p_{B}&=&p_{A}+(p_{a}+p_{B})+\frac{1}{\kappa}p_{2}-\frac{1}{\kappa}p_{v}.
\end{eqnarray}
As a result of Alice's measurement the operators $x_{u}$ and $p_{v}$ collapse in modes 
$a$ and $B$ into the classical variables $\bar x_{u}$ and $\bar p_{v}$. After the measurement
Alice and Bob perform on the modes the displacements 
\begin{eqnarray}\label{displacements}
x_{a,{\rm out}}&=&x_{a}'-\frac{1}{\kappa}\bar x_{u},\quad 
p_{a,{\rm out}}=p_{a}'-\frac{1}{\kappa}\bar p_{v},\nonumber\\
x_{B,{\rm out}}&=&x_{B}-\frac{1}{\kappa}\bar x_{u},\quad
p_{B,{\rm out}}=p_{B}+\frac{1}{\kappa}\bar p_{v},
\end{eqnarray}
thus obtaining the following output quadratures:
\begin{eqnarray}\label{output}
x_{a,{\rm out}}&=&x_{A}-\frac{1}{\kappa}x_{1}-\kappa x_{2},\nonumber\\ 
p_{a,{\rm out}}&=&-p_{A}-\kappa p_{1}-\frac{1}{\kappa} p_{2},\nonumber\\
x_{B,{\rm out}}&=&x_{A}-(x_{a}-x_{B})-\frac{1}{\kappa}x_{1},\nonumber\\
p_{B,{\rm out}}&=&p_{A}+(p_{a}+p_{B})+\frac{1}{\kappa}p_{2}.
\end{eqnarray}
The fidelities $F_{A}$ and $F_{B}$ between the input coherent state and the output states
$\rho_{A}$ and $\rho_{B}$ can be calculated from the formula
$F_{i}=1/(1+\langle n_{\rm ch,i}\rangle)$, where $\langle n_{\rm ch,i}\rangle$ is the mean 
number of chaotic photons in $i$th mode. Hence, one finally arrives
using the Eqs.~(\ref{cA}) and (\ref{output}) at the following formulas:
\begin{equation}\label{CVapprox}
F_{A}=\frac{2}{2+e^{2\gamma}},\quad F_{B}=\frac{2}{2(1+e^{-2r})+e^{-2\gamma}}, 
\end{equation}
where $\gamma=\ln\kappa$. Comparing the obtained fidelities with those of given in Eq.~(\ref{asymF}), 
we find that while the operation fidelity $F_{A}$ attains optimal value the teleportation 
fidelity $F_{B}$ is reduced in comparison with the optimal value (\ref{asymF}) 
due to the finiteness of the shared entanglement. Therefore, in contrast with the 
qubit protocol our CV teleportation scheme realizes only approximately optimal PQT 
of coherent states. The teleportation fidelity $F_{B}$ can approach the optimum (\ref{asymF}) 
arbitrarily closely if sufficiently large CV entanglement (sufficiently large squeezing $r$) 
is available.
\section{Conclusion}\label{sec_conclusion}

In conclusion, we have explicitly constructed a joint two-qubit non-demolition measurement allowing  
to partially deterministically discriminate among all Bell states. Moreover, this partial non-demolition 
Bell measurement allows to continuously control the degree of the discrimination 
by the preparation of the state of ancilla. We have then proposed the 
teleportation scheme based on the partial Bell measurement making it 
possible to continuously control the flow of information between the output qubits and      
it was proved that such a distribution of information is optimal in the sense that the 
teleportation and operation fidelity saturate the cloning inequality
(\ref{inequality}). Further, the measurement was shown to be optimal two-qubit quantum 
operation from the point of view of the trade-off between the gain of information and the state 
disturbance. Analogic partial non-demolition Bell measurement can be constructed for 
CVs. However, in contrast with qubits the CV partial quantum teleportation based on 
the measurement does not distribute the information optimally while the optimum is 
approached with increasing shared entanglement.
\medskip
\section*{Acknowledgments}
 
We would like to thank P. Marek for stimulating discussions. 
The research has been supported by Project LN00A015 and Research
Project No. CEZ: J14/98 of the Czech Ministry of Education. 
R. F. acknowledges support by Project 202/03/D239 of the 
Grant Agency of the Czech Republic.

\end{document}